\documentclass[twoside,twocolumn]{article}
\usepackage[super,sort&compress,comma]{natbib}
\usepackage{mhchem}
\usepackage{times,mathptmx}
\usepackage{sectsty}
\usepackage{balance}
\usepackage{color}

\usepackage{multirow}
\usepackage{bm}% bold math
\usepackage[mathlines]{lineno}

\usepackage{graphicx} %eps figures can be used instead
\usepackage{subfigure}
\usepackage{amsmath}
\usepackage{lastpage}
\usepackage{fancyhdr}
\usepackage{hyperref}

\graphicspath{{figures/}{../figures/}}

\begin{document}

\thispagestyle{plain}
\fancypagestyle{plain}{
\fancyhead[L]{\includegraphics[height=8pt]{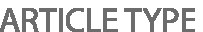}}
\fancyhead[C]{\hspace{-1cm}\includegraphics[height=20pt]{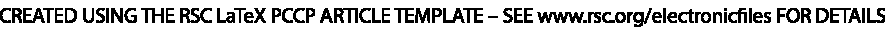}}
\fancyhead[R]{\includegraphics[height=10pt]{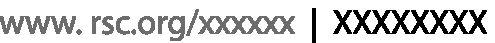}\vspace{-0.2cm}}
\renewcommand{\headrulewidth}{1pt}}
\renewcommand{\thefootnote}{\fnsymbol{footnote}}
\renewcommand\footnoterule{\vspace*{1pt}%
\hrule width 3.4in height 0.4pt \vspace*{5pt}}
\setcounter{secnumdepth}{5}

\makeatletter
\def\hlinew#1{%
  \noalign{\ifnum0=`}\fi\hrule \@height #1 \futurelet
   \reserved@a\@xhline}
\makeatother

\newcommand*{\citen}[1]{%
  \begingroup
    \romannumeral-`\x % remove space at the beginning of \setcitestyle
    \setcitestyle{numbers}%
    \cite{#1}%
  \endgroup
}

\makeatletter
\def\subsubsection{\@startsection{subsubsection}{3}{10pt}{-1.25ex plus -1ex minus -.1ex}{0ex plus 0ex}{\normalsize\bf}}
\def\paragraph{\@startsection{paragraph}{4}{10pt}{-1.25ex plus -1ex minus -.1ex}{0ex plus 0ex}{\normalsize\textit}}
\renewcommand\@biblabel[1]{#1}
\renewcommand\@makefntext[1]%
{\noindent\makebox[0pt][r]{\@thefnmark\,}#1}
\makeatother
\renewcommand{\figurename}{\small{Fig.}~}
\sectionfont{\large}
\subsectionfont{\normalsize}

\fancyfoot{}
\fancyfoot[LO,RE]{\vspace{-7pt}\includegraphics[height=9pt]{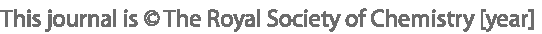}}
\fancyfoot[CO]{\vspace{-7.2pt}\hspace{12.2cm}\includegraphics{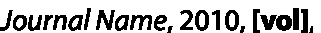}}
\fancyfoot[CE]{\vspace{-7.5pt}\hspace{-13.5cm}\includegraphics{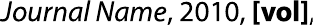}}
\fancyfoot[RO]{\footnotesize{\sffamily{1--\pageref{LastPage} ~\textbar  \hspace{2pt}\thepage}}}
\fancyfoot[LE]{\footnotesize{\sffamily{\thepage~\textbar\hspace{3.45cm} 1--\pageref{LastPage}}}}
\fancyhead{}
\renewcommand{\headrulewidth}{1pt}
\renewcommand{\footrulewidth}{1pt}
\setlength{\arrayrulewidth}{1pt}
\setlength{\columnsep}{6.5mm}
\setlength\bibsep{1pt}

\twocolumn[
  \begin{@twocolumnfalse}

\noindent\LARGE{\textbf{Concurrent coupling of atomistic simulation and mesoscopic hydrodynamics for flows over soft multi-functional surfaces}}
\vspace{0.6cm}

\noindent\large{\textbf{Yuying Wang,\textit{$^{a,b,c,*}$} Zhen Li,\textit{$^{b,*,\dag}$} Junbo Xu,\textit{$^{a}$} Chao Yang,\textit{$^{a,c,\dag}$} and George Em Karniadakis \textit{$^{b}$}}}\vspace{0.5cm}
%Please note that \ast indicates the corresponding author(s) but no footnote text is required.

\noindent\textit{\small{\textbf{Received Xth XXXXXXXXXX 2018, Accepted Xth XXXXXXXXX 2018
%\newline First published on the web Xth XXXXXXXXXX 201X
}}}

%\noindent \textbf{\small{DOI: 10.1039/b000000x}}
\vspace{0.6cm}
\noindent \normalsize{We develop an efficient parallel multiscale method that bridges the atomistic and mesoscale regimes, from nanometer to micron and beyond, via concurrent coupling of atomistic simulation and mesoscopic dynamics. In particular, we combine an all-atom molecular dynamics (MD) description for specific atomistic details in the vicinity of the functional surface, with a dissipative particle dynamics (DPD) approach that captures mesoscopic hydrodynamics in the domain away from the functional surface. In order to achieve a seamless transition in dynamic properties we endow the MD simulation with a DPD thermostat, which is validated against experimental results by modeling water at different temperatures. We then validate the MD-DPD coupling method for transient Couette and Poiseuille flows, demonstrating that the concurrent MD-DPD coupling can resolve accurately the continuum-based analytical solutions.
Subsequently, we simulate shear flows over polydimethylsiloxane (PDMS)-grafted surfaces (polymer brushes) for various grafting densities, and investigate the slip flow as a function of the shear stress. We verify that a ``universal" power law exists for the sliplength, in agreement with published results.
Having validated the MD-DPD coupling method, we simulate time-dependent flows past an endothelial glycocalyx layer (EGL) in a microchannel. Coupled simulation results elucidate the dynamics of EGL changing from an equilibrium state to a compressed state under shear by aligning the molecular structures along the shear direction. MD-DPD simulation results agree well with results of a single MD simulation, but with the former more than two orders of magnitude faster than the latter for system sizes above one micron.}
\vspace{0.5cm}
\end{@twocolumnfalse}
]

\section{Introduction}
\footnotetext{$^*$ The first two authors contribute equally to this work.}
\footnotetext{\textit{$^{a}$CAS Key Laboratory of Green Process and Engineering, Institute of Process Engineering, Chinese Academy of Sciences, Beijing 100190, China}}
\footnotetext{\textit{$^{b}$Division of Applied Mathematics, Brown University, Providence, RI 02912, USA}}
\footnotetext{\textit{$^{c}$University of Chinese Academy of Sciences, Beijing 100049, China}}
\footnotetext{$^\dag$ Correspondence: zhen\_li@brown.edu, chaoyang@ipe.ac.cn}

Using tethered chains or brush-like layers we can design effective soft multi-functional surfaces for diverse engineering applications. For instance, tethered polymer chains or brushes in Micro-Electro-Mechanical systems (MEMS) or membrane technologies have attracted great attention due to their potential for flow regulation~\cite{2002saito,2007wang,2017chen}. As another example, the combination of traditional ceramic membranes and organic polymers yields diverse surface properties, extending its usage to a wide range of solvents~\cite{2011Vargas}. In addition to regulating the wettability, the polymer grafting can realize pore size tuning at the same time, which makes ceramic membranes more versatile~\cite{2015Tanardi}. However, grafting polymers also brings complexity to the flow system and therefore it is more difficult to analyze it or improve its design. A traditional approach is to perform different types of analysis for different flow subsystems. For example, the classic diffusion laws can be used in the bulk region while models for hindered flow in porous media can be employed for the polymer region~\cite{2000Castro}. Sometimes, this two-region analysis can be replaced by a boundary modification to the regular models~\cite{2015Tanardi}, but the boundary conditions are not easy to obtain due to the complex interactions between the polymers and flow.
\\\indent
On the biological side, one particular example is the endothelial glycocalyx layer (EGL), coating the endothelial cells and lining entire vascular system~\cite{2000Pries}. Glycocalyx is a sugar-rich layer, formed by oligosaccharide chains in direct contact with blood~\cite{2014Alphonsus}. Recognized as an immobile sheet in early times, the EGL was subsequently found to interfere with the ambient flow and dramatically increase the microvascular flow resistance~\cite{1990Desjardins,1994Pries}. The EGL is in direct contact with blood and essential to human metabolism, as in their interaction with blood, they act as protective layer for chemical or mechanical irritations. Moreover, few biological process that happen in the vessels can avoid the participation of EGL. Therefore, their dendritic structure and their dynamics under flow and interactions with blood cells or medicaded particles are all of great interest. Recent efforts have focused on understanding EGL along with growing concerns about health problems such as diabetes~\cite{2007Perrin} and atherosclerosis~\cite{2007Reitsma}. Though experimental technologies can be used to probe the flow-glycocalyx interaction, we can gain a deeper insight into the EGL-blood interaction using appropriate computational models.
\\\indent
Among the existing simulation techniques, molecular dynamics (MD) is the most suitable method for modeling polymer brushes and EGL. Recent studies~\cite{2018Jiang,2014CruzChu,2018Pikoula} have demonstrated the effectiveness of MD in EGL modeling. However, simulating a vascular system is computationally expensive, and in fact to conduct an all-atom molecular simulation is prohibitively expensive and perhaps unnecessary, since EGL occupies only the endothelial regions of the vessels. Nevertheless, the complex interaction between blood flow and glycocalyx requires extremely high resolution, while the surrounding flow field has to be resolved adequately at a much larger spatio-temporal scales. We encounter the same problem in modeling polymer-brushes, where a detailed description maybe necessary to capture the complex dynamics, and we require multiple simulations to form a complete understanding since the flow depends on several factors such as grafting density, polymer length and shear stress~\cite{2009Yong,2015Speyer}. For both the polymer brush and EGL problems, a single-scale simulation can be either computationally prohibitive or too coarse-grained to capture the important physics. Multiscale modeling approaches can employ heterogeneous descriptions, e.g., continuum and atomistic, hence combining different computational and resolution advantages. Several examples of multiscale methods are included in references~\citen{2004Nie,2005Werder,2007Dupuis,2009Fedosov,2014Olson,2015Bian_PRE}.

In the multiscale modeling, a variety of models is employed to have different levels of resolution and complexity to study one system, where these models are coupled either analytically or numerically. Conceptually, multiscale methods for coupling different solvers can be classified into two categories, namely sequential multiscale modeling and concurrent multiscale modeling~\cite{2011E}. The sequential coupling approach couples a hierarchy of computational models by sequentially transferring information, so that large-scale models can use the information obtained from more detailed small-scale models~\cite{2018Zhao}. For this reason, the sequential coupling approach is also referred to as serial message-passing method. Alternatively, the concurrent multiscale modeling considers the quantities at each scale depends strongly on what happens at the other scales, so that different computational models are coupled on-the-fly as the computation proceeds. In a concurrent multiscale simulation, domain decomposition is often used to partition the system into sub-domains characterized by different scales and physics, and then the different scales of the system are coupled concurrently by a hand-shaking procedure~\cite{2018Bian_book}. Examples include a hybrid simulation coupling a lattice Boltzmann solution of the Navier-Stokes equations to a MD simulation of a dense fluid~\cite{2007Dupuis}, and a triple-decker algorithm applied to atomistic-mesoscopic-continuum simulations of shear flows~\cite{2009Fedosov}. In the present study, we consider soft multi-functional surfaces subject to time-dependent shear flows with focus on the concurrent multiscale modeling implemented with the domain decomposition method.

We employ dissipative particle dynamics (DPD) as our mesoscopic model to describe coarse-grained dynamics that is coupled with atomistic dynamics. Same as the MD model, DPD is a particle-based simulation method modeling stochastic dynamics associated with correct fluctuation correlations. Because current fluctuations can play an important role in microscale dynamics in molecular systems, without consideration of mesoscale fluctuations and correlations, some important physical features beyond the mean-field theory predictions, i.e., attraction of similarly charged plates~\cite{2002Lau}, may not be able to modeled. Moreover, DPD using a Lagrangian description of molecular systems has a direct connection with the MD method, because the governing equations of DPD can be rigorously derived by applying the Mori-Zwanzig projection to an atomistic dynamics~\cite{2014Li}. In addition, the mean-field hydrodynamic equations of a DPD system recover the Navier-Stokes equations in the continuum limit~\cite{1997Marsh}. Therefore, DPD is a good candidate to be coupled to the MD solver in our multisccale problems as it can seamlessly bridge the nanoscale dynamics and the mesoscale hydrodynamics.

The remainder of this paper is organized as follows: in Section~\ref{sec:2} we briefly introduce the atomistic and mesoscopic methods, as well as the details how to implement the concurrent coupling via domain decomposition by matching state variables. In Section~\ref{sec:3}, we first validate the multiscale coupling method for transient Couette and Poiseuille flows. We subsequently investigate polydimethylsiloxane-grafted surfaces with various grafting densities subjected to shear flows, and then simulate time-dependent flows past an endothelial glycocalyx layer in a microchannel. Finally, we end up with a brief summary and discussion in Section~\ref{sec:4}.

\section{Mathematical models}\label{sec:2}
\begin{figure}[t!]
\centering
\includegraphics[width=0.46\textwidth]{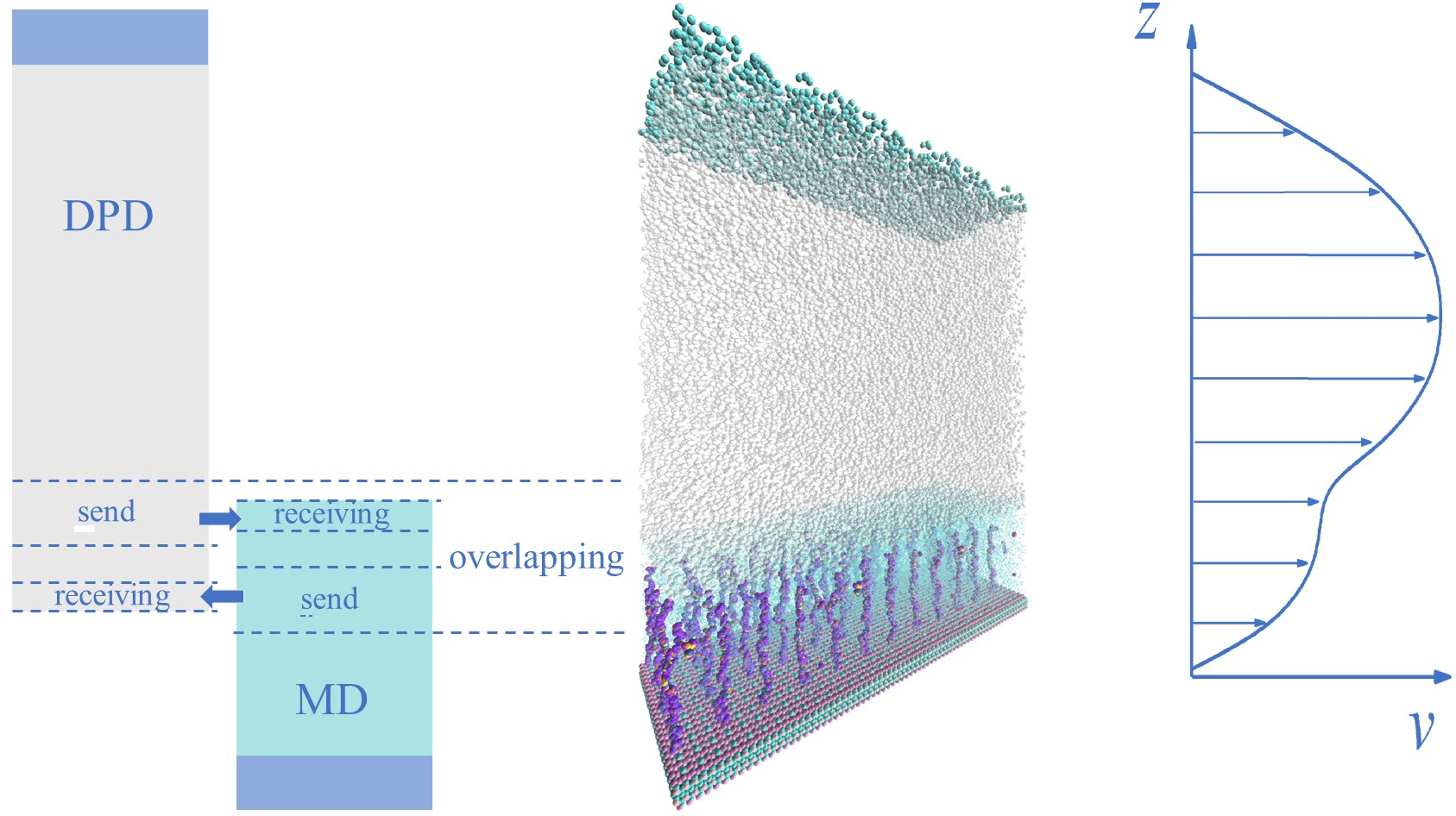}
\caption{Schematic of the domain decomposition of a three-dimensional channel (middle), and the MD-DPD coupling via an overlapping domain (left). On the right, a typical flow profile is sketched. Shown in the MD domain are polymer brushes and solvent whereas in DPD domain we include only the solvent (middle). The MD domain is much smaller than the DPD domain, but the MD simulation is computationally much more expensive than the DPD simulation.}
\label{fig:sketch}
\end{figure}

The function of soft functional surfaces depends on the surface nanostructures with designed physical/chemical features, which are, in general, determined by specific atomistic details. Therefore, we employ an all-atom MD description to capture the molecular details in the subdomain with a functionalized surface, while we use a coarse-grained approach by DPD in the outer domain to simulate the flow of solvent and possibly the transport of nanoparticles. As shown in Fig.~\ref{fig:sketch}, the MD and DPD systems are then coupled via a domain decomposition technique by matching the state variables in an overlapping region. In this section, we first briefly introduce the atomistic and mesoscopic methods, and then describe in detail how to implement the concurrent coupling via domain decomposition.

\subsection{Atomistic model}
\begin{figure}[t!]
\centering
\includegraphics[width=0.9\columnwidth]{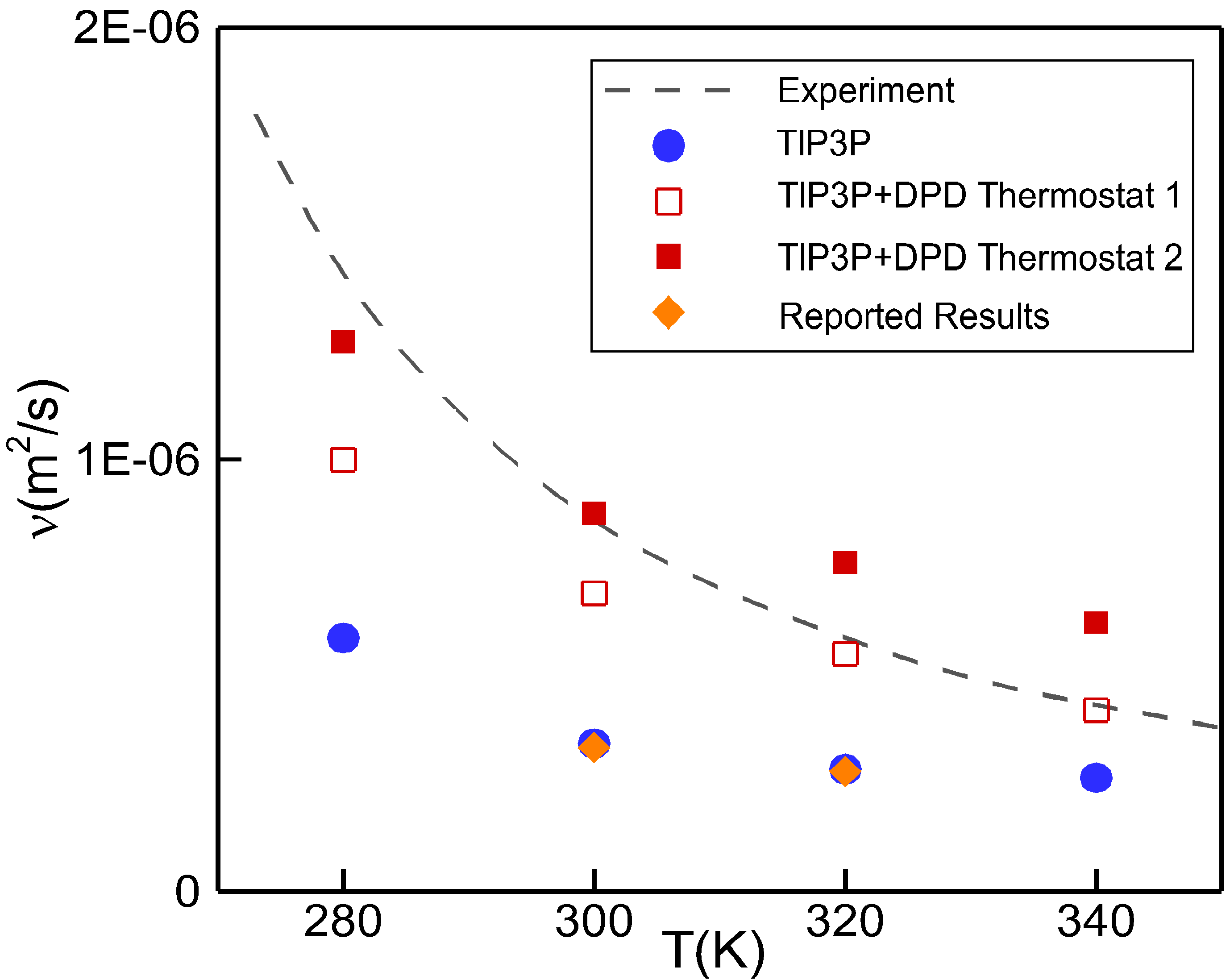}
\caption{Dependence of kinematic viscosity of water on temperature. Our MD simulation results from the TIP3P model (solid circles) are compared with TIP3P results at $300~K$ and $320~K$ by Mao and Zhang~\cite{2012Mao} (diamonds, orange). Results of TIP3P with DPD thermostats with $r_{c}=0.25~nm$ are shown by open squares and with $r_{c}=0.3~nm$ by solid squares. The dashed line shows experimental results~\cite{crchandbook}.}
\label{fig:DPDthermo}
\end{figure}

To test the capability and limitations of the concurrent coupling algorithm, we will consider three molecular systems, namely a simple fluid system, a polymer grafted system, and a glycocalyx system. The simple fluid system is made of water molecules, which also constitute the solvent in the other two systems. A Lenard-Jones potential is applied among all atoms, wherein the TIP3P model is used for water. Previous works have demonstrated the poor performance of classic water models in representing water transport properties such as viscosity~\cite{2010Gonz,2010Yanmei}. For example, our simulation with the traditional TIP3P model yields very low kinematic viscosity as shown in Fig.~\ref{fig:DPDthermo}, which is in agreement with previous reports~\cite{2012Mao,2010Yanmei}. Such low values of viscosity may lead to an erroneous velocity field in response to particular shear stress. An effective way to correct the water viscosity is to employ the DPD thermostat in the MD system~\cite{2007Junghans}. The DPD thermostat, which consists of the pair of dissipative force and the random force, not only serves as a thermostat but also regulates the viscosity of the fluid. Given the dissipative force in the form of $\mathbf{F}^D_{ij}=-\gamma(1-r_{ij}/r_c)^s(\mathbf{e}_{ij}\mathbf{v}_{ij})\mathbf{e}_{ij}$, the kinematic viscosity can be roughly estimated by a function that depends on the DPD parameters as~\cite{2014Lienergy}
\begin{equation}\label{eq:nu}
\begin{split}
 \nu =&\frac{3k_{B}T(s+1)(s+2)(s+3)}{16 \pi\gamma\rho_n r_{c}^{3}} \\
 &+\frac{16 \pi\gamma\rho_n r_{c}^{5}}{5(s+1)(s+2)(s+3)(s+4)(s+5)},
\end{split}
\end{equation}
where $\rho_n$ is the number density of particles. Eq.~\eqref{eq:nu} implies that the kinematic viscosity can be regulated by the dissipative parameter $\gamma$, cut-off radius $r_{c}$ and the exponent of weight function $s$. In the MD system, we fix $s=0.5$ and $\gamma=5.56\times10^{-15}~Nm/s$ to test the regulation effect on viscosity by changing $r_{c}$. The simulation results are plotted in Fig.~\ref{fig:DPDthermo} for $r_{c}=0.25~nm$ and $r_{c}=0.3~nm$. We observe an increase in the viscosity values compared to the classical TIP3P model, and the DPD thermostat with $r_{c}=0.3~nm$ yields correct kinematic viscosity $\nu=8.90\times10^{-7}~m^2/s$ at $300~K$ that is consistent with experimental measurements.

In order to model {\it polymer brushes}, we employ polydimethylsiloxane~(PDMS), a polymer used widely in industrial applications. It combines characteristics such as flexibility, thermal/chemical stability and hydrophobicity, and has been used as grafted polymer in previous works~\cite{2014Pinheiro,2014Tanardi,2015Tanardi}. In our work, we employ PDMS as the grafted polymers and silica as the substrate to study the polymer response under water flow. The PDMS is represented using the atomistic model in the MD solver. The Lenard-Jones 12-6 potential is applied to the atoms, where the atomistic united atom force-field is used for the PDMS polymers. This force field has been validated against experiments, and accurate radial distribution functions were obtained~\cite{2007Muakrodimitri}. Also, the aforementioned TIP3P water model with DPD thermostat is used for water. The PDMS/water interaction was validated by the simulation of a water droplet on PDMS matrix, with a contact angle of~$104^{\circ}$ in our work and in agreement with~$100^{\circ}$-$110^{\circ}$ from previous experiments~\cite{2006Bodas}.

In order to model the {\it glycocalyx}, we adopt the most detailed all-atom model so far introduced by Cruz-Chu et al.~\cite{2014CruzChu} In this model, the sugar chains are represented by heparan sulfate (HS) chains as they are the most prevalent oligosaccharides in glycocalyx.\ Also, Syndecan-4 (Syn-4) are chosen as the transmembrane proteins embedded in a palmitoyl-oleoyl phosphatidylcholine (POPC) lipid bilayer. We employ the CHARMM force field to compute the atomistic interactions between the glycocalyx atoms and the protein atoms~\cite{1998Mackerell}.

\subsection{Mesoscopic model}
A mesoscopic model covers the region where atomistic details can be neglected in order to reduce the computational cost. The DPD method~\cite{1998Groot,2017Espanol} is one of the most popular mesoscopic models, with governing equations rigorously derived by applying the Mori-Zwanzig projection to a MD system~\cite{2014Li}. In this study, we adopt DPD as our mesoscopic model to describe coarse-grained dynamics that is coupled with atomistic dynamics.

Similarly to the MD model, a DPD system consists of many interacting particles, each of them is considered as a coarse-grained particle representing the collective dynamics of a group of $N_{c}$ molecules. The pairwise interactions between DPD particles are governed by three forces, namely the conservative force $\mathbf{F}_{ij}^{C}$, the dissipative force $\mathbf{F}_{ij}^{D}$, and the random force $\mathbf{F}_{ij}^{R}$. The formulas for these forces are given below:
\begin{equation}
\begin{split}
    \mathbf{F}^{C}_{ij}&=a_{ij}\omega_{C}(r_{ij})\textbf{\emph{e}}_{ij},\\
    \mathbf{F}^{D}_{ij}&=-\gamma_{ij}\omega_{D}(r_{ij})(\textbf{\emph{e}}_{ij}\cdot\textbf{\emph{v}}_{ij})\textbf{\emph{e}}_{ij},\\
    \mathbf{F}^{R}_{ij}&=\delta_{ij}\omega_{R}(r_{ij})\xi_{ij}\Delta t^{-1/2}\textbf{\emph{e}}_{ij},
\end{split}
\end{equation}
where $a$, $\gamma$ and $\epsilon$ denote the strengths of the forces, $r_{ij}$ is the distance between particles $i$ and $j$, $\mathbf{e}_{ij}$ is the unit vector from particle $j$ to $i$, and $\mathbf{v}_{ij} = \mathbf{v}_{i} - \mathbf{v}_{j}$ is the velocity
difference. Also,~$\omega_{C}(r),~\omega_{D}(r)$ and $\omega_{R}(r)$ are the weight functions of $\mathbf{F}^{C}$, $\mathbf{F}^{D}$ and $\mathbf{F}^{R}$, respectively. We choose the weight functions as $\omega_{C}(r) = (1-r/r_c)$ and $\omega_{D}(r) = (1-r/r_c)^{0.5}$ for $r\leq r_c$ with a cutoff radius $r_c$, beyond which the weight functions vanish. The fluctuation-dissipative theory requires $\omega_D(r)=\omega_R^2(r)$ and $\delta^2 = 2\gamma k_BT$, where $k_B$ is the Boltzmann constant and $T$ the temperature.

To impose correctly the no-slip boundary condition on the channel walls and prevent density fluctuations, we employ a corrected dissipative coefficient as introduced by Li et al.~\cite{2018Li} This correction is a function of the distance between the particle and the wall and it is only activated when the distance is smaller than the cut-off radius $r_{c}$.

\subsection{Concurrent coupling}
{\bf Matching of state variables:}
The coupling is achieved by state exchange, i.e., the exchange of particle velocities in the overlapping region. As shown in Fig.~\ref{fig:sketch}, the velocity information from the sending regions of both domains are saved for use. During each communication, the velocity of the DPD particles lying within the receiving region is set as the average velocity of the MD particles in the same area. Similarly, the MD particles in the receiving region acquire the updated velocities in the same way. Spatial averaging with the Gaussian kernel is performed when extracting the local velocity. At the outer border of the overlapping region, both the MD and DPD systems impose a particle reflection to prevent losing particles. Any particle locating beyond the outer border is moved back to the computational domain by a specular reflection, and its velocity is then updated by assigning the velocity obtained from other solver. Before each communication, the MD simulation runs for 10 steps while the DPD simulation runs for one step. Therefore, we can obtain smooth velocity profile like the one shown in Fig.~\ref{fig:sketch}.

Both MD and DPD simulations are performed using the Large-scale Atomic/Molecular Massively Parallel Simulator (LAMMPS)~\cite{1995Plimpton}. The concurrent coupling of the MD and DPD solvers is implemented with the MUI library~\cite{2015Tang}. MUI is a C++ header-only library and serves as the data exchange and interpretation layer between different solvers, so that data of state variables at discrete points can be easily passed in a push-fetch manner. The three dimensional rendering of the MD and DPD systems is generated using VMD~\cite{1996Humphrey}.

\noindent
{\bf Matching of physical properties:}
The MD system describes a physical system at the atomistic level, while the DPD system is a coarse-grained representation of a physical system. The typical length and time scales in a MD simulation are in general different from those in a DPD simulation. Correct coupling of MD and DPD system requires that the MD and DPD systems have consistent physical properties so that they correspond to the same physical system. To this end, we need to first perform a unit mapping between the MD system and the DPD system. Let the lower case symbols $[l]$, $[m]$ and $[t]$ be the three basic units (i.e., length, mass and time units) of the MD system, and the capital case symbols $[L]$, $[M]$ and $[T]$ be the length, mass and time units of the DPD system, respectively. Given a coarse-graining level $N_C$, a unit mapping from the MD system to the DPD system can be determined.

Taking the liquid water at temperature of $300~K$ as an example, $[m]=2.99\times10^{-26}~kg$ is the mass of a water molecule, $[l]=3.17\times10^{-10}~m$ is the characteristic length of a TIP3P water molecule, and $[t]=1\times10^{-15}~s$ is the time step used in MD simulations. Given a coarse-graining level $N_c=10$, the mass of a DPD particle is $[M]=N_c\cdot[m]=2.99\times10^{-25}~kg$. Then, the length scale of the DPD system can be computed by matching the physical mass density. Let $\rho_n=3.0$ be the number density of DPD particles, the mass density of the DPD liquid should be $\rho = \rho_n\cdot[M]/[L]^3 = 1000~kg/m^3$, from which we have $[L]=9.64\times10^{-10}~m$. Let $[\epsilon]$ and $[\nu]$ be the units of energy and kinematic viscosity of the DPD system. Although the time unit $[L]$ can be determined via a dimensional analysis of either energy $[\epsilon]$ or viscosity $[\nu]$, i.e., $[\epsilon]=[M][L]^2[T]^{-2}$, and $[\nu]=[L]^2[T]^{-1}$, the time units computed from matching the energy and matching the viscosity should be consistent so that correct magnitude of thermal fluctuations can be captured. Because both the reduced DPD temperature and the kinematic viscosity of DPD fluids change with the value of energy unit $[\epsilon]$, we vary the value of $[\epsilon]$ carefully and compute the corresponding viscosity until we have a consistent time unit from the energy and the viscosity.
By running a few test DPD systems, we obtain the energy unit $[\epsilon]=9.78\times10^{-21}~J$ and the viscosity unit $[\nu]=1.74\times10^{-7}~m^2/s$, both leading to a time unit $[T]=5.33\times10^{-12}~s$. Accordingly, we have a DPD system with $k_BT = 0.42$ and $\nu = 5.11$ in reduced DPD units representing liquid water with a kinematic viscosity of $8.90\times10^{-7}~m^2/s$ at a temperature of $300~K$. For comparison, the values of the characteristic units of both MD and DPD models are listed in Table~\ref{tab:unit}.

\setlength{\tabcolsep}{4pt}
\begin{table}[t!]
\centering
\begin{tabular}{lccc}
\hline\hline
\multirow{2}{*}{} & MD                  & DPD                  & Reduced    \\
                  & (real units)        & Scales               & DPD units  \\
\hline
Length ($m$)    &$3.17\times10^{-10}$  &$9.64\times10^{-10}$   & 1                  \\
CG level        &-                     &10                     & -                  \\
Mass ($kg$)     &$2.99\times10^{-26}$  &$2.99\times10^{-25}$   & 1                  \\
Time ($s$)      &$1\times10^{-15}$     &$5.33\times10^{-12}$   & 1                  \\
Energy ($J$)    &$4.14\times10^{-21}$  &$9.78\times10^{-21}$    & 0.42              \\
Viscosity ($m^2/s$) & $8.90\times10^{-7}$ &$1.74\times10^{-7}$  &$5.11$             \\
%Velocity ($m/s$)&$1\times10^{5}$       &181                    \\
%Force ($N$)     &$6.95\times10^{-11}$  &$1.02\times10^{-11}$   \\
\hline\hline
\end{tabular}
\caption{\label{tab:unit} Matching the physical properties: units for MD and DPD simulations. We perform MD simulations using LAMMPS with real units, and perform DPD simulations using LAMMPS with reduced DPD units. In coupled MD-DPD simulations, both MD and DPD quantities are converted to physical units to matching state variables}
\end{table}

\section{Results}\label{sec:3}
\subsection{Simple shear flows}
\begin{figure}[t!]
\centering
\includegraphics[width=0.98\columnwidth]{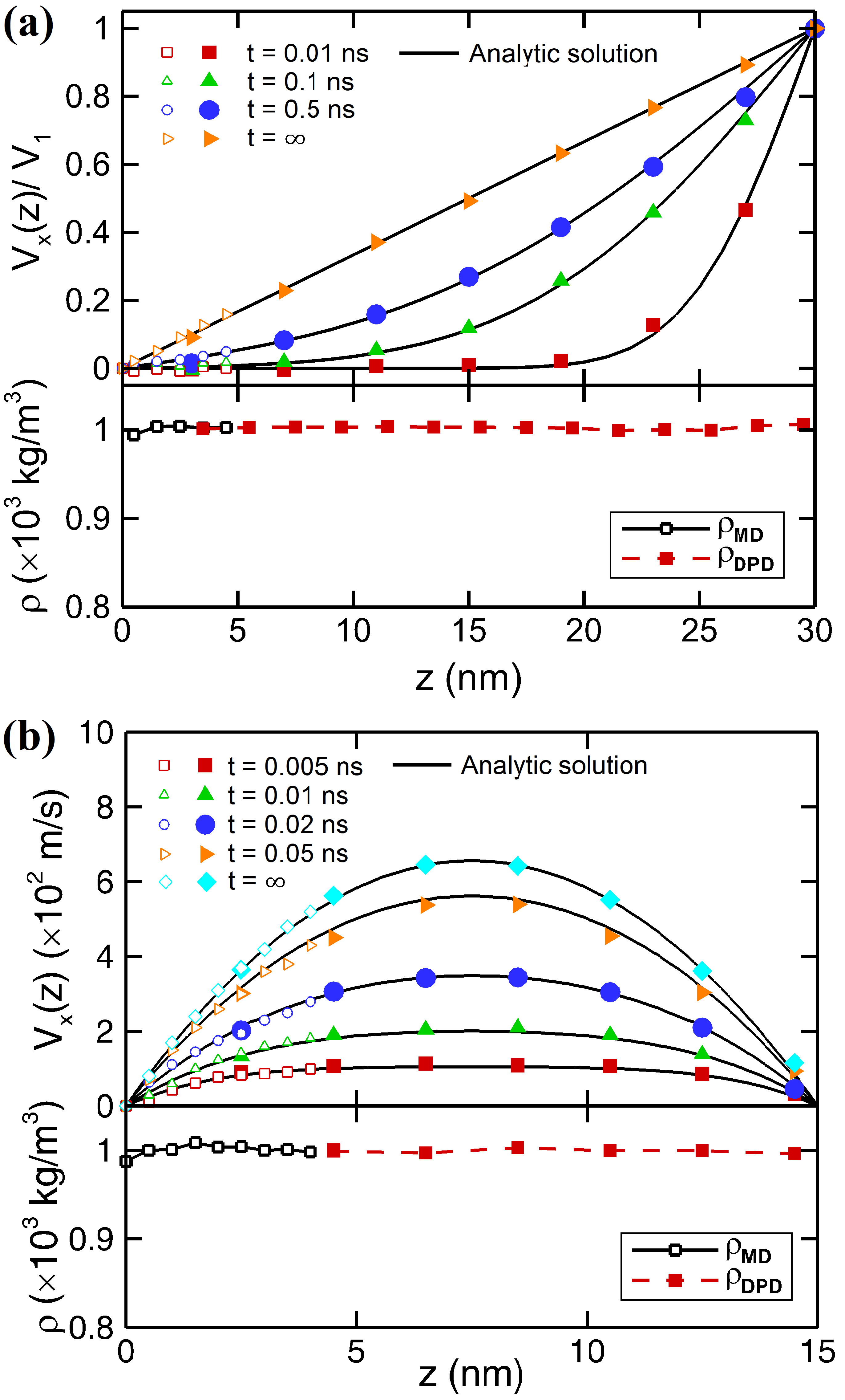}
\caption{Testing the MD-DPD coupling for simple shear flows. Shown are time-dependent velocity profiles and density profiles for (a) Couette flow, and (b) Poiseuille flow. The overlapping domain is $2.5~nm$. The solid symbols represent DPD results while the open symbols represent MD results.}
\label{fig:test}
\end{figure}

To validate the MD-DPD coupled algorithm, we firstly apply it to the cases of simple flows, namely Couette flow and Poiseuille flow. The arrangement of the MD and DPD solvers is based on the overlapping domains shown in Fig.~\ref{fig:sketch}. For the {\it Couette flow}, the fluid is confined between two parallel walls at a distance of $30~nm$.  The MD domain extends from $0$ to $5~nm$ while the DPD domain extends from $2.5$ to $30~nm$ with an overlapping region of $2.5~nm$. Both domains are periodic in $x$ and $y$ directions. We set the wall velocity in $x$-direction as $v_{0}=0.0$ on the MD side, and $v_{1}=1.0$ in reduced DPD unit on the DPD side. The evolution of velocity profiles with time is computed and shown in Fig.~\ref{fig:test}(a).

In the case of {\it Poiseuille flow}, the distance between walls is $15~nm$. The MD domain extends from $0$ to $4~nm$ from the lower wall while the DPD domain extends from $2.5$ to $15~nm$ with an overlapping region of $1.5~nm$. Periodic boundary conditions are applied in $x$ and $y$ directions in both solvers. Let $m_{a}$ be the atom mass and $m_{w}$ be the mass of a water molecule in MD units. The walls are fixed on both sides, while body forces at $0.0005~m_{a}$ in the MD domain and $0.0005~N_{c}~m_{w}[F_{\rm MD}]/[F_{\rm DPD}]$ in the DPD domain are applied to the flow particles. Here, $[F_{\rm MD}]$ and $[F_{\rm DPD}]$ represent the force units in MD and DPD solvers, respectively. The same as for the Couette flow, we compute the time-evolution of velocities profiles of the Poiseuille flow shown in Fig.~\ref{fig:test}(b).

In the Couette flow, the time evolution of velocity profile can be described analytically as~\cite{2015Li_tDPD}
\begin{equation}
V_{x}(z,t)=\frac{U_0z}{d} + \sum_{n=1}^{\infty}\frac{2U_0}{n\pi}(-1)^n\exp(-\nu\varphi_n^2 t)\sin(\varphi_n z),
\end{equation}
where $\varphi_n = n\pi/d$ with $d$ being the channel width, $U_0$ is the velocity of the moving wall, $\nu$ is the kinematic viscosity and $t$ the time.

On the other hand, the time-dependent Poiseuille flow follows the analytic solution~\cite{2014Lienergy}
\begin{equation}
\begin{split}
V_{x}(z,t)=&\frac{gd^{2}}{8\nu}\left[1-\left(\frac{2z}{d}\right)^{2}\right]
-\sum_{n=0}^{\infty} \frac{4(-1)^{n}d^{2}}{\nu\pi^{3}(2n+1)^{3}}\cdot \\ &\cos\left[\frac{(2n+1)\pi z}{d}\right]\exp\left(-\frac{(2n+1)^{2}\pi^{2}\nu t}{d^{2}}\right),
\end{split}
\end{equation}
where $g$ is the body force.
Fig.~\ref{fig:test} shows the comparison between the coupled simulations and the analytical solutions. We observe excellent agreement in both Couette and Poiseuille flow. In addition, we observe an approximately uniform density across the domains. A slightly drop of the density at the wall is due to stiff atomistic interaction, while constant density profile is observed in the bulk MD region and in the DPD domain. Here, it is worthy noting that, in order to eliminate the density fluctuations in the vicinity of wall due to the atomistic lattice structures, we create the solid wall by cutting selected regions from a thermal equilibrium MD system so that the particle distribution in the solid phase is randomized and has the same particle structures as the liquid phase~\cite{2018Li}. Benchmarks of both the Couette and Poiseuille flows demonstrate that the coupled MD-DPD method provides continuous and smooth velocity profiles across the overlapping region and can represent accurate time-dependent flows.

\subsection{Polymer brushes}
\begin{figure} \centering
\includegraphics[width=1.0\columnwidth]{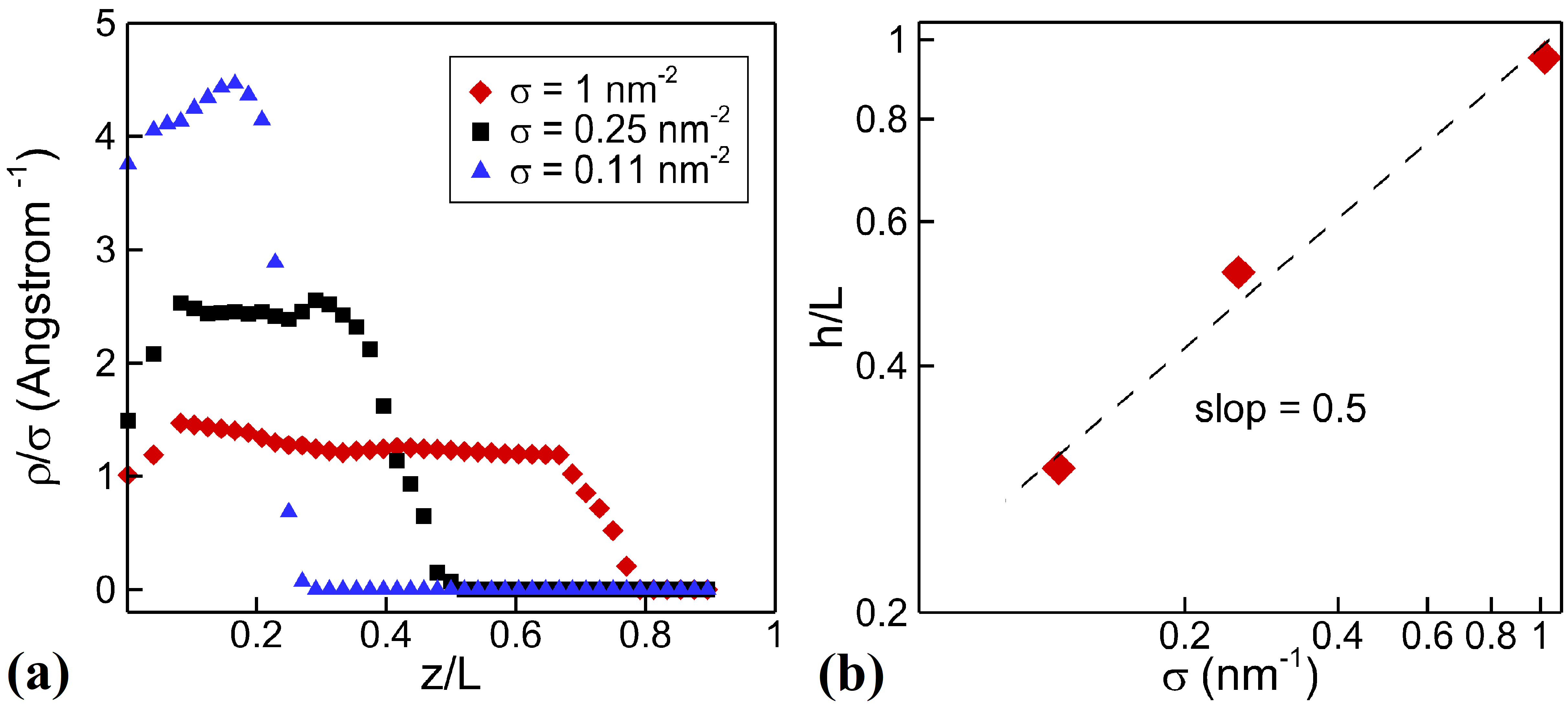}
\caption{Polymer brushes in equilibrium. (a) polymer density as a function of the distance from the wall for different grafting densities. (b) polymer height as a function of grafting density. The MD simulation results match the theoretical slope of 0.5~\cite{2012Deng,1999Kent,1988Milner}.}
\label{fig:polymerdensity}
\end{figure}

To validate the PDMS polymer model, we first perform equilibrium simulations only with the MD solver. According to some widely accepted results, the equilibrium properties of polymer brushes are determined by the excluded-volume interaction and conformation entropy of the polymer chains, and the polymer heights follows the scaling law as~\cite{2012Deng,1999Kent,1988Milner}:
\begin{equation}
h \sim L(\sigma b^{2})^{\beta},
\end{equation}
where $\sigma$ is the grafting density,~$L$ is the contour length of polymer, and $h$ is the average brush height, which can be calculated from:
\begin{equation}
h=2\frac{\int z\rho(z)dz}{\int \rho(z)dz}.
\end{equation}
The exponent $\beta$ tends to be different in good solvent and theta solvent. Due to the hydrophobicity of PDMS, there should be $\beta=1/2$ for the PDMS and water system~\cite{2012Deng}.
We perform the PDMS-grafted simulations for three different grafting densities; from dense to sparse, we have $\sigma = 1.0$, $0.25$, and $0.11~nm^{-2}$, respectively. In Fig.~\ref{fig:polymerdensity} we show results of the brush height, which are in agreement with the theoretical scaling law.

\begin{figure}
\centering
\includegraphics[width=0.95\columnwidth]{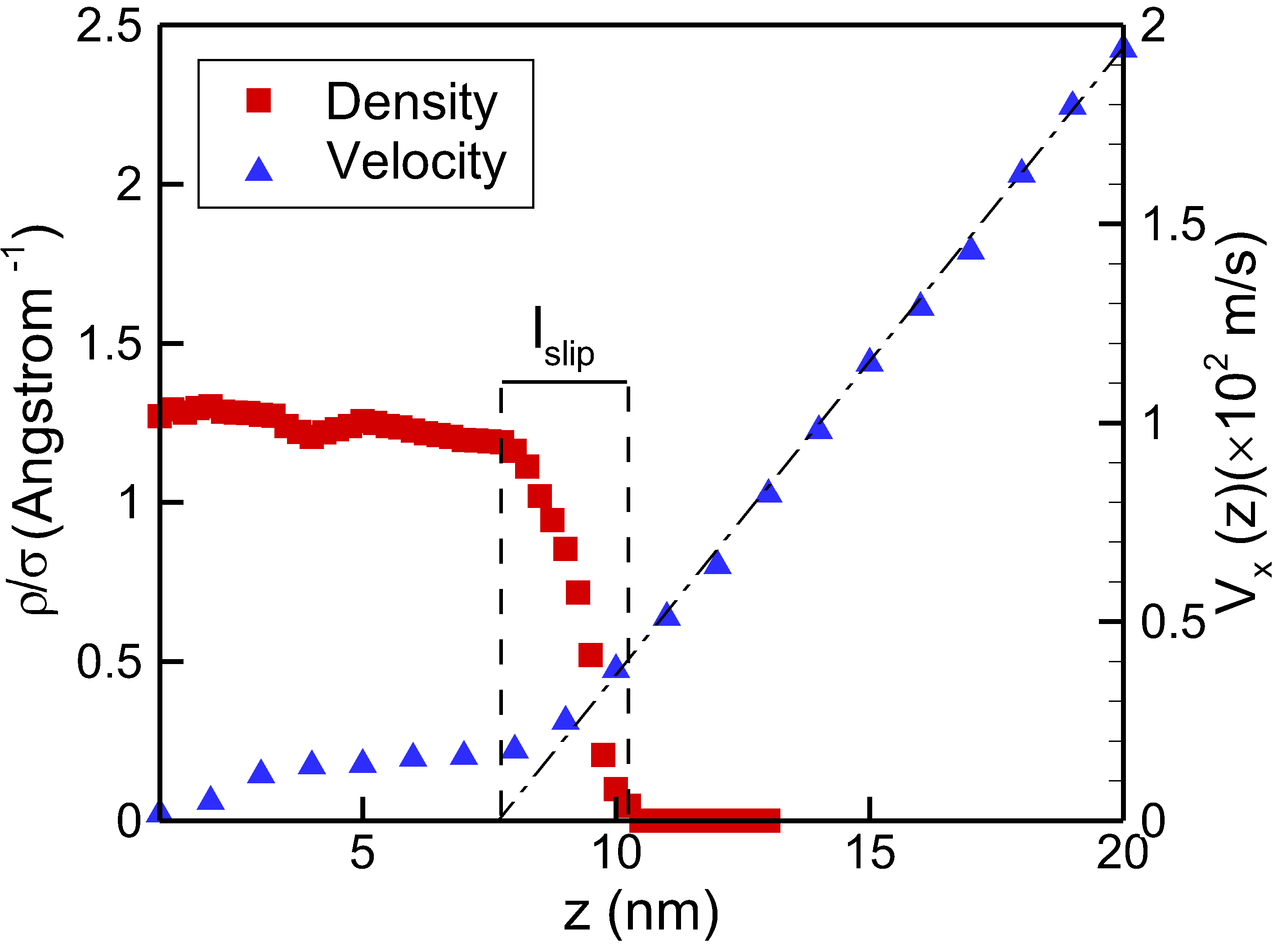}
\caption{Polymer brushes subject to shear flow. Shown are density and velocity profiles over a PDMS-grafted surface for grafting density $\sigma=1~nm^{-2}$. The definition of slip length is also illustrated in the figure.}
\label{fig:definesliplength}
\end{figure}

We drive the flow by imposing a fixed velocity at the upper wall on the DPD side. Fig.~\ref{fig:definesliplength} shows the velocity profile for water and the density profile for polymer, from which we can see that the flow inside the brush is greatly affected compared to the standard Couette flow, shown a plug-like profile. By extrapolating the far-field linear velocity profile to zero, we can find an imaginary boundary site that has the same far-field velocity profile by replacing the brush with a solid wall at a new position. The penetration depth, also called slip length, is defined from the tip-position of polymers to the extrapolating site as shown in Fig.~\ref{fig:definesliplength}, where the tip-position is determined by the location where the density decay to zero for polymer brushes in equilibrium. Thus, the polymer grafted wall can be approximatively regarded as a solid wall with the thickness of $(h-l_{\rm slip})$.

\begin{figure}
\centering
\includegraphics[width=1.0\columnwidth]{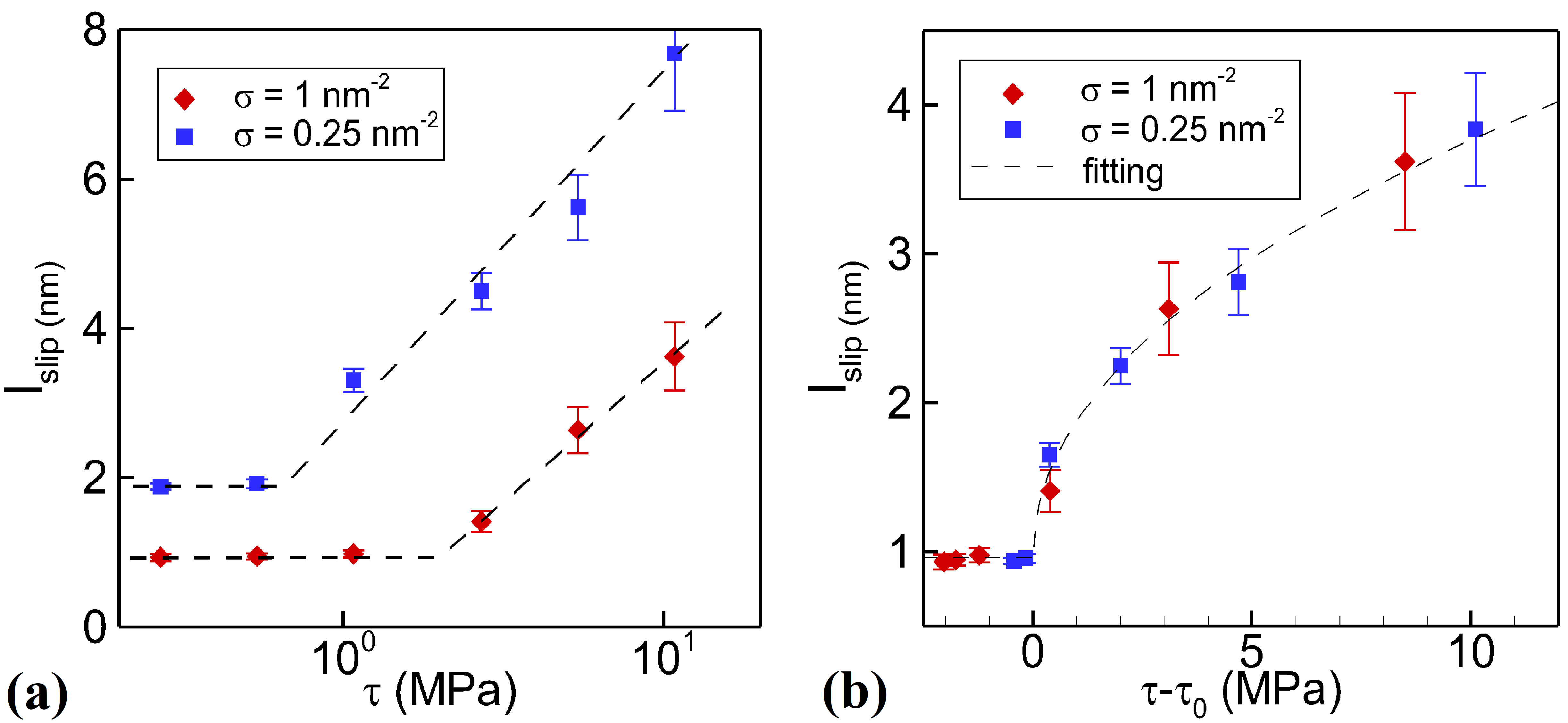}
\caption{Polymer brushes subject to shear flow. (a) slip length versus shear stress for two different grafting densities.  (b) Slip length replotted versus the difference of shear stress and the critical shear stress $\tau_{0}$. Results collapse to a universal curve as predicted by Eqs.~\eqref{con:scalling1}-\eqref{con:scalling3}.}
\label{fig:scall}
\end{figure}

The slip length depends on both the grafting density and the shear rate; it increases with shear rate, partly due to polymer bending under high shear stress and partly due to the high velocity in the near wall region with almost flat velocity profile. In Fig.~\ref{fig:scall}(a) we display the variation of the slip length with shear rate in different systems with various grafting densities.  Increasing the shear rate can compress the polymer and also enhance the hindered flow in the boundary regions, leading to larger slip length.
Let $\tau$ be the average shear stress defined by the dynamical viscosity multiplying the velocity gradient in the bulk region. We find in Fig.~\ref{fig:scall}(a) that the slip length remains almost unchanged when $\tau$ is relatively small. However, as $\tau$ increases beyond a critical value $\tau_{0}$, $l_{\rm slip}$ starts increasing linearly. The critical shear stress $\tau_0$ is different for various grafting densities. It is appreciated that as in denser grafted systems, stronger stresses are required to compress the polymers, hence expanding the region of the bulk flow. In addition, the hindered flow in the polymer domain also requires greater shear stress values to form a visible velocity with higher grating density. In summary, higher grafting density can lead to higher critical shear stress for the slip length to increase with the stress.

In the previous scaling law based on the so-called ``blob theory", we have $l_{\rm slip}\sim \sigma^{-1/2}$ for specific kinds of polymers~\cite{1990Rabin}. Deng et al.~\cite{2012Deng} demonstrated that such scaling law can only be used below the critical shear rate, with the overall scaling law for the slip length changing with shear rate given as
\begin{equation}\label{con:scalling1}
   \frac{l_{\rm slip}}{\sigma^{-1/2}}\sim 1+f(\tau-\tau_{0},\beta) ,
\end{equation}
where $f$ is a step-like function as
\begin{equation}
f(\tau-\tau_{0},\beta)=\left\{
\begin{array}{ccl}
0 && {\tau<\tau_{0}}\\
(\tau-\tau_{0})^{\alpha(\beta)} && {\tau\geq\tau_{0}}. \label{con:scalling2}
\end{array} \right.
\end{equation}
Therefore, if we plot $l_{\rm slip}/\sigma^{-1/2}$ with $(\tau-\tau_{0})$, there should be a unique curve for a specific type of polymer in a specific solvent, regardless of the grafting density. So, we re-plot the data in Fig.~\ref{fig:scall}(a) in the form of $(l_{\rm slip}/\sigma^{-1/2})$ with $(\tau-\tau_{0})$ for different $\sigma$, and indeed they collapse to a single curve as shown in Fig.~\ref{fig:scall}(b). By fitting this curve, we obtain an expression for $f(\tau-\tau_{0})$ as follows

\begin{equation}
f(\tau-\tau_{0},\beta)=\left\{
\begin{array}{ccl}
0 && {\tau<\tau_{0}}\\
(\tau-\tau_{0})^{0.5} && {\tau\geq\tau_{0}}\label{con:scalling3}.
\end{array} \right.
\end{equation}

\subsection{Glycocalyx}
\begin{figure}[t!]
\centering
\includegraphics[width=1.0\columnwidth]{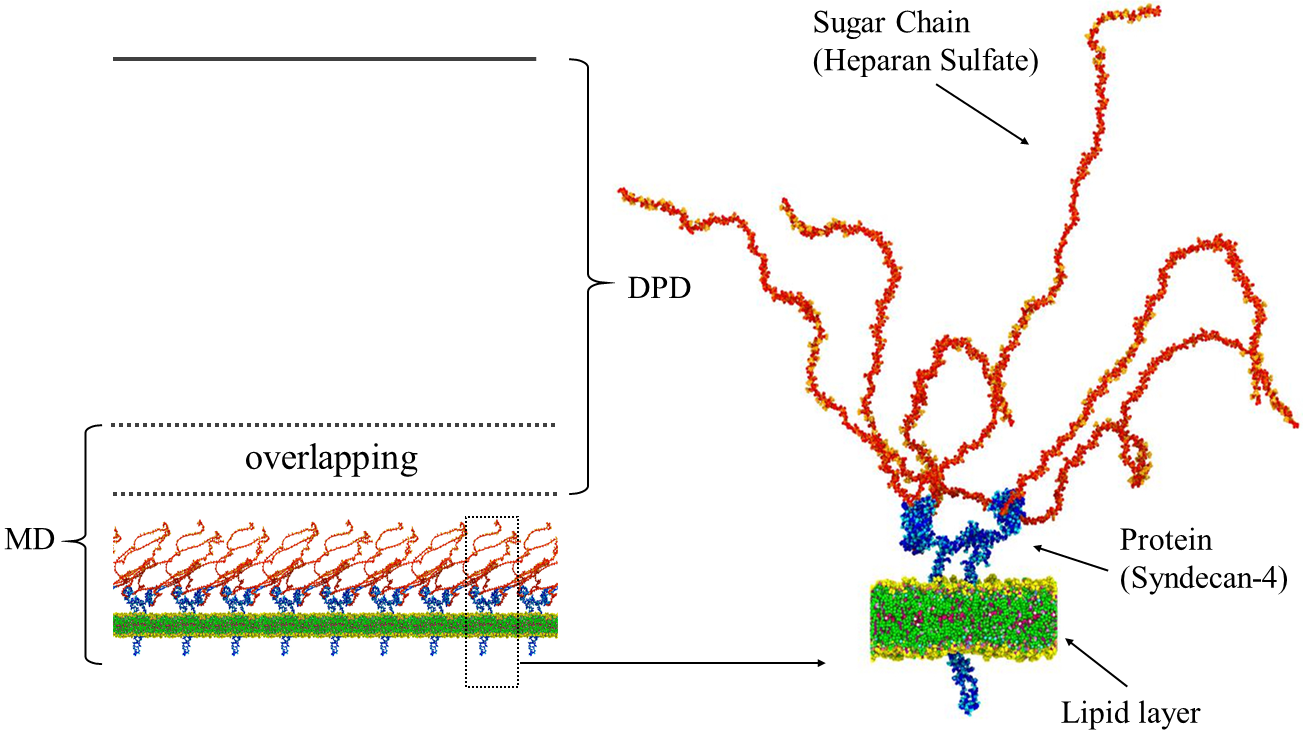}
\caption{\label{fig:sketchglyco} Schematic of the domain decomposition for the MD-DPD simulation of glycocalyx in a 3D channel subject to shear flow (left). On the right the detailed structure of the glycocalyx based on Cruz-Chu et al.'s work~\cite{2014CruzChu} is shown.}
\end{figure}

The all-atom structure of glycocalyx first introduced by Cruz-Chu et al.~\cite{2014CruzChu} is adopted in our simulations. As shown in Fig.~\ref{fig:sketchglyco}, each glycocalyx unit contains a Syndecan-4 dimer, and each of the monomers is joined with three 50-residue heparan sulfate sugar chains. We start with a small system with one glycocalyx unit and $50~nm$ height in $z$-direction beyond the lipid layer. We perform simulations both with the MD model and the coupled MD-DPD model. Subsequently, we enlarge the domain to a height of $200~nm$ and also $1~\mu m$ to compare the computational efficiency between the MD and the MD-DPD algorithms.

\begin{figure}[t!]
\centering
\includegraphics[width=0.8\columnwidth]{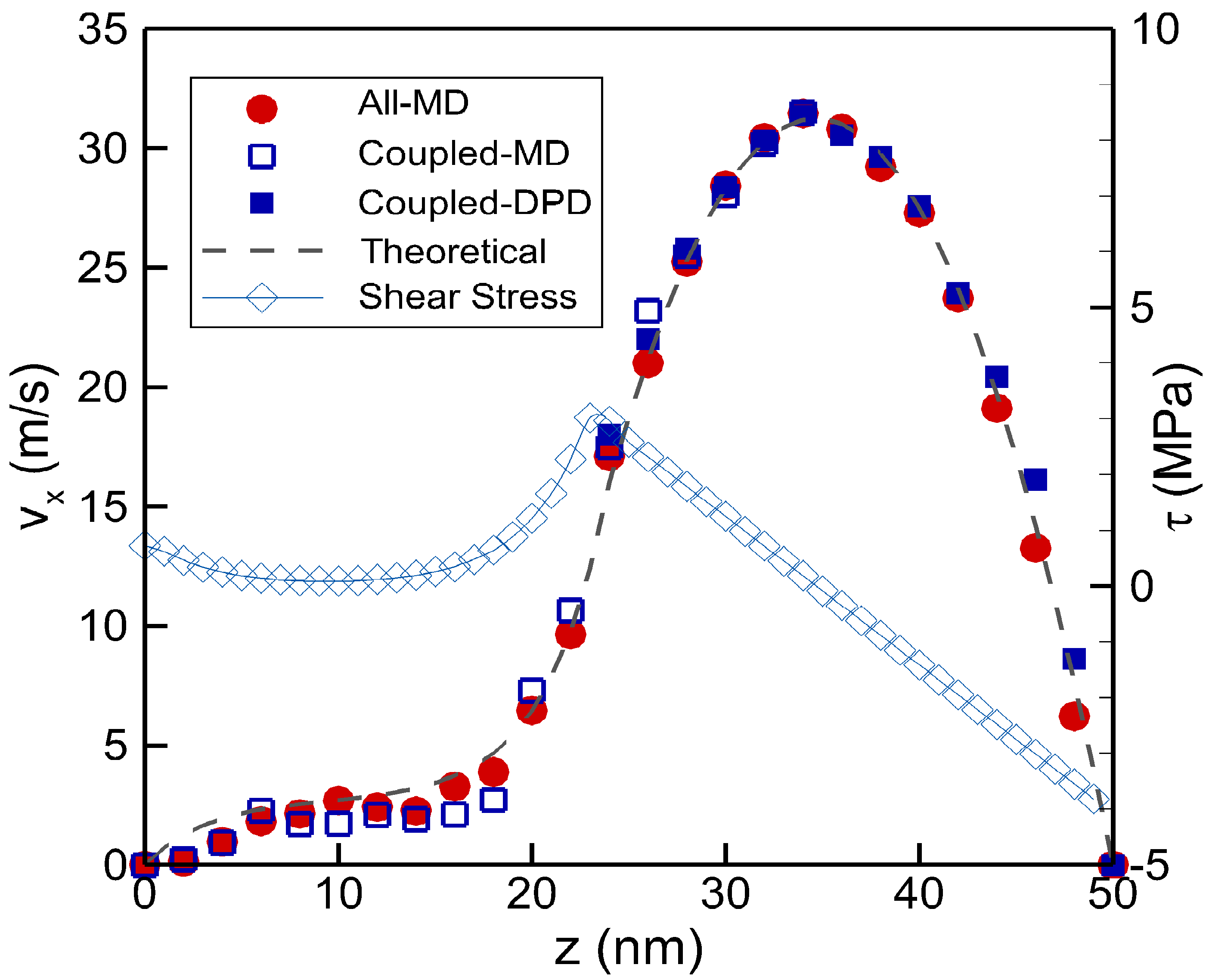}
\caption{Velocity profiles and shear stress (open diamond) over the glycocalyx layer and across the channel. Velocities are obtained with all-atomic MD simulation (red circles), MD-DPD coupled simulation (open and solid squares), and continuum based numerical results based on the model of Eq.~\eqref{con:brinkman} (dash line). The overlapping domain is $10~nm$.}
\label{fig:allMD}
\end{figure}

In Fig.~\ref{fig:allMD} we display the velocity profile of the pressure-driven flow over a single glycocalyx model when the upper boundary wall is $50~nm$ away from the lipid layer. For runs with the MD simulation alone, the velocity profiles are denoted by circles. In accordance with previous works~\cite{2014CruzChu}, a parabolic velocity profile occurs above the glycocalyx, while the profile seems to be linear within the sugar chains. Subsequently, we employ the coupled MD-DPD system to repeat the same simulations. Here the MD domain extends from $0$ to $35~nm$ while the DPD domain extends from $25$ to $50~nm$ with an overlapping region of $10~nm$. The DPD domain contains only water molecules. The velocity profile obtained from the coupled MD-DPD simulation, displayed as squares, is in good agreement with the MD simulation.

We can also introduce some modeling based on the continuum hypothesis in order to derive simple solutions of the glycocalyx system. To this end, the pressure-driven flow through the glycocalyx layer can be described by the Brinkman equation~\cite{1949Brinkman,2003Weinbaum}, which is used for permeation through porous membranes
\begin{equation}\label{con:brinkman}
  \mu\frac{d^{2} v_{x}}{dz^{2}}-\Lambda v_x+\frac{dP}{dx}=0,
\end{equation}
where $v_{x}$ is the fluid velocity in the $x$-direction and $P$ is the pressure; $\mu$ denotes the dynamic viscosity, and $\Lambda$ is the hydraulic resistivity, which depends on Darcy's permeability $\kappa$ by $\Lambda=\mu/\kappa$.

Beyond the glycocalyx layer, there is a fully developed flow, where the momentum equation is
\begin{equation}\label{con:stokes}
  \mu\frac{d^{2} v_{x}}{dz^{2}}+\frac{dP}{dx}=0.
\end{equation}
We employ a second-order finite difference method to obtain a numerical solution for Eqs.~\eqref{con:brinkman} and \eqref{con:stokes}, where $\mu=0.9\times 10^{-3}~Pa\cdot s$ and $\nabla P=2.4\times10^{8}~MPa/m$. The value of $\Lambda$ varies within the range between $9\times 10^{9}$ and $9\times10^{11}~dyn\cdot s/cm^{4}$. We recorded the mean squared errors when fitting the numerical solution to the velocity plots from our MD-DPD coupled simulation, and the best fitting comes from the $\Lambda=9\times 10^{10}~dyn\cdot s/cm^{4}$ case, which lies exactly within the range given by previous estimates~\cite{2007Weinbaum} of $10^{10}$-$10^{11}~dyn\cdot s/cm^{4}$. The numerical solution when $\Lambda=9\times 10^{10}~dyn\cdot s/cm^{4}$ is also displayed in Fig.~\ref{fig:allMD}.

\begin{figure}[t!]
\centering
\includegraphics[width=1.0\columnwidth]{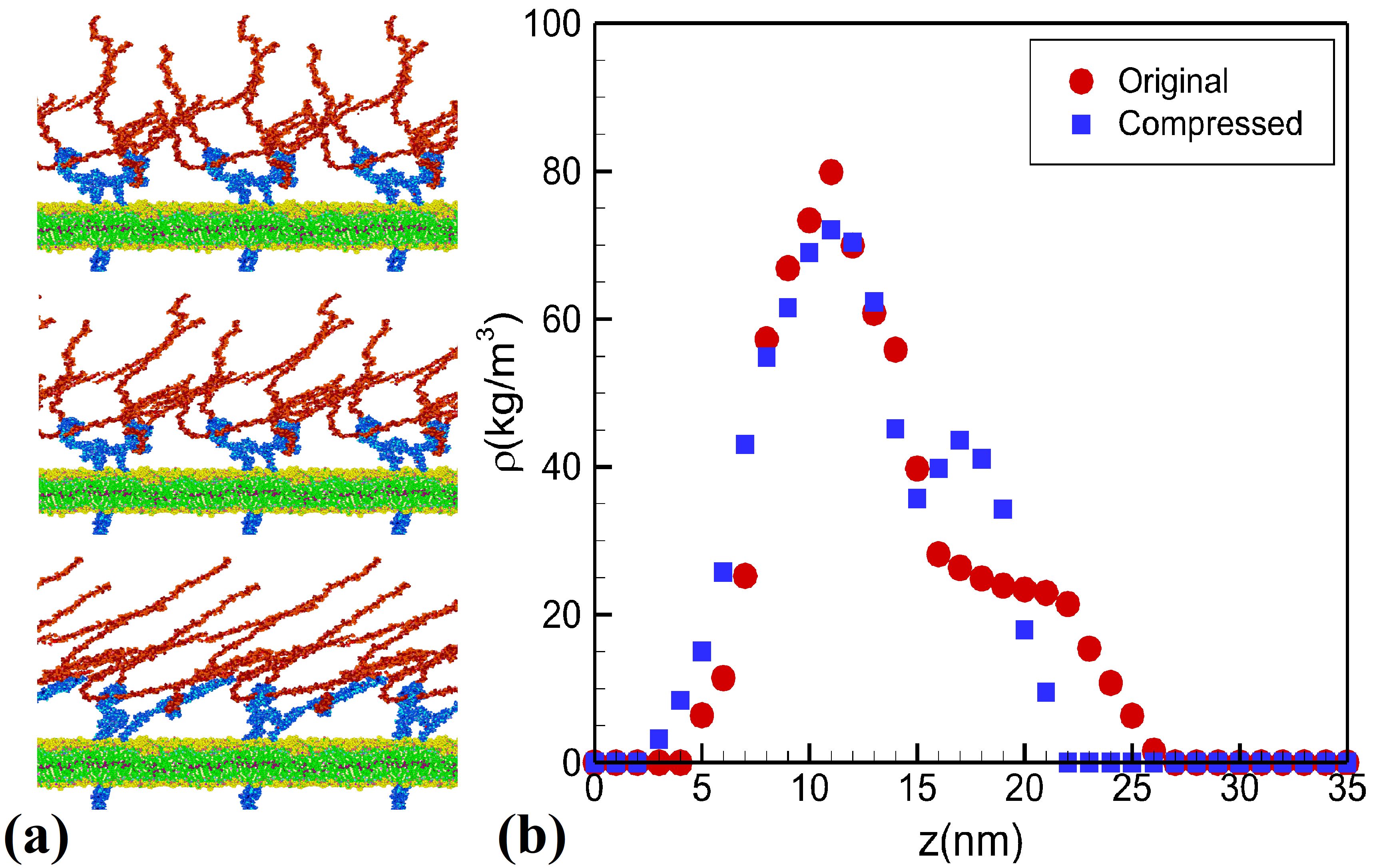}
\caption{Simulation results of glycocalyx subject to shear flow.~(a) Snapshots of the glycocalyx and the lipid layer at initial equilibrium (top), at intermediate stretching (middle), and fully stretched (bottom). (b) Sugar chain density at initial configuration (circles) and in fully stretched configuration (squares). See also Supporting Information for a movie.}
\label{fig:gly}
\end{figure}

With the MD-DPD coupled algorithm, we can enlarge the system by duplicating the glycocalyx layer three times. In the simulation we observe the compression of the glycocalyx layer under flow as shown in Fig.~\ref{fig:gly} (see also Supporting Information for a movie). The density profiles of the sugar chains before and after the compression are also displayed in Fig.~\ref{fig:gly}. When comparing the density profile with the velocity in Fig.~\ref{fig:allMD} we can find a mobile region above $15~nm$ from the wall and a relatively stable region below that. From the shear stress profiles in Fig.~\ref{fig:allMD} we can obtain similar results, i.e., that the maximum shear stress occurs at the tip-area of the sugar chains while their lower parts are subject to much smaller stresses that do not affect the glycocalyx structure.

\begin{figure}[t!]
\centering
\includegraphics[width=0.8\columnwidth]{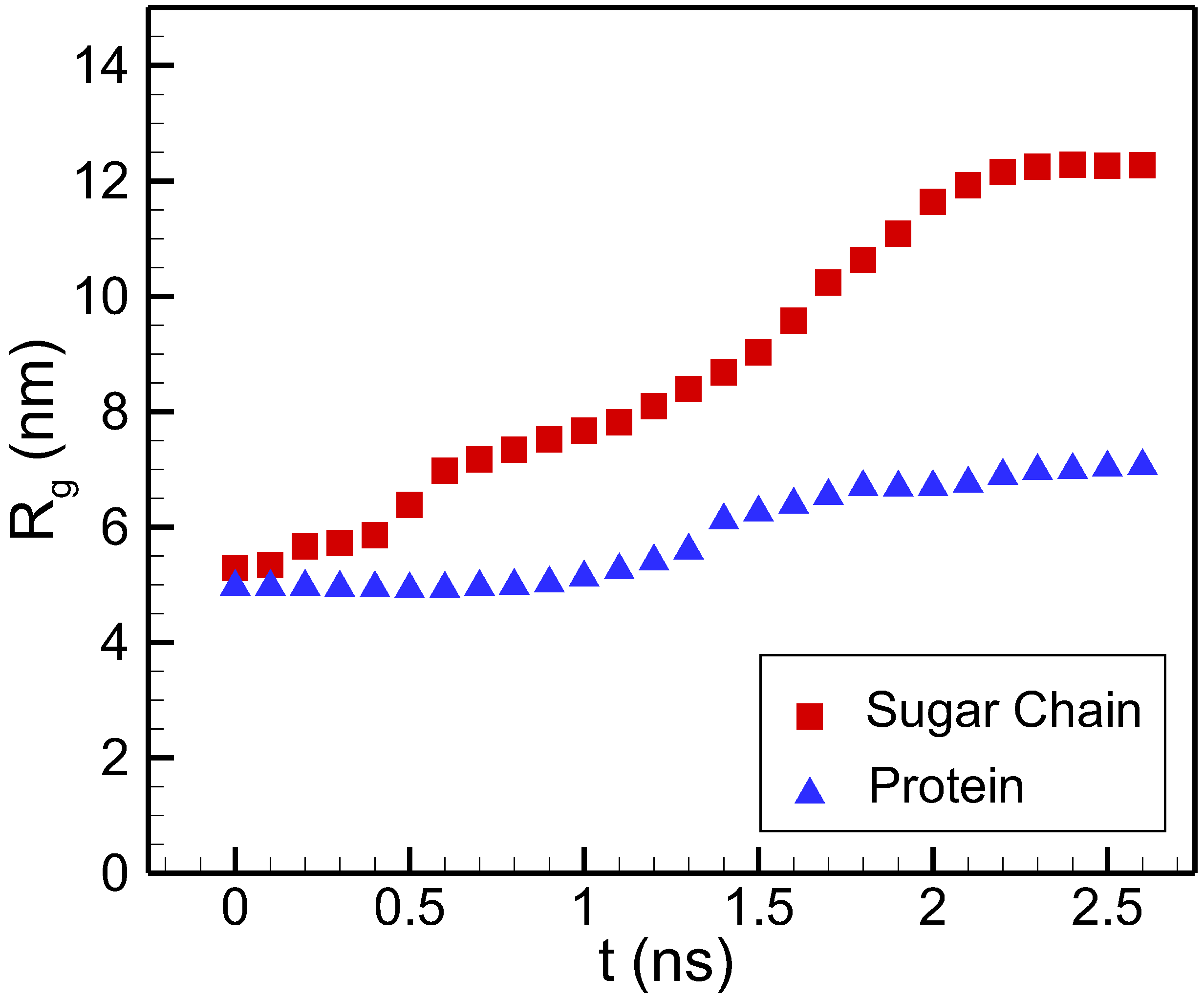}
\caption{Evolution of the radius of gyration under shear flow. The square symbols denote the sugar chain and the triangle symbols denote the protein. Steady state is achieved after about $2~ns$. }
\label{fig:gyration}
\end{figure}

To further quantify the glycocalyx distortion under flow, we use gyration tensor to represent its three-dimensional structure~\cite{2012Lei}. The gyration tensor is defined as
\begin{equation}
G_{mn}=\frac{1}{N_{v}}\sum_{i}(r_{m}^{i}-r_{m}^{c})(r_{n}^{i}-r_{n}^{c}),
\end{equation}
where $m$,~$n$ can be $x$,~$y$ or $z$, and $r^{i}$ denotes the atomic positions while $r^{c}$ is the center of mass. If we denote the eigenvalues of the gyration tensor as $\lambda_{1}$, $\lambda_{2}$ and $\lambda_{3}$, then the radius of gyration can be given as $R_{g}=(\lambda_{1}+\lambda_{2}+\lambda_{3})^{1/2}$.
In Fig.~\ref{fig:gyration} we display the evolution of $R_{g}$ with time for both the sugar chains and proteins. The time starts after the formation of a steady flow, and the results for the sugar chains represent the average value of all six chains on the central monomer.
As we see from Fig.~\ref{fig:gyration}, the gyration radius of both sugar chains and protein increases with time. This demonstrates that they are both stretched by the flow, and after some time the glycocalyx layer reaches statistically stationary state.

\begin{figure}[t!]
\centering
\includegraphics[width=0.8\columnwidth]{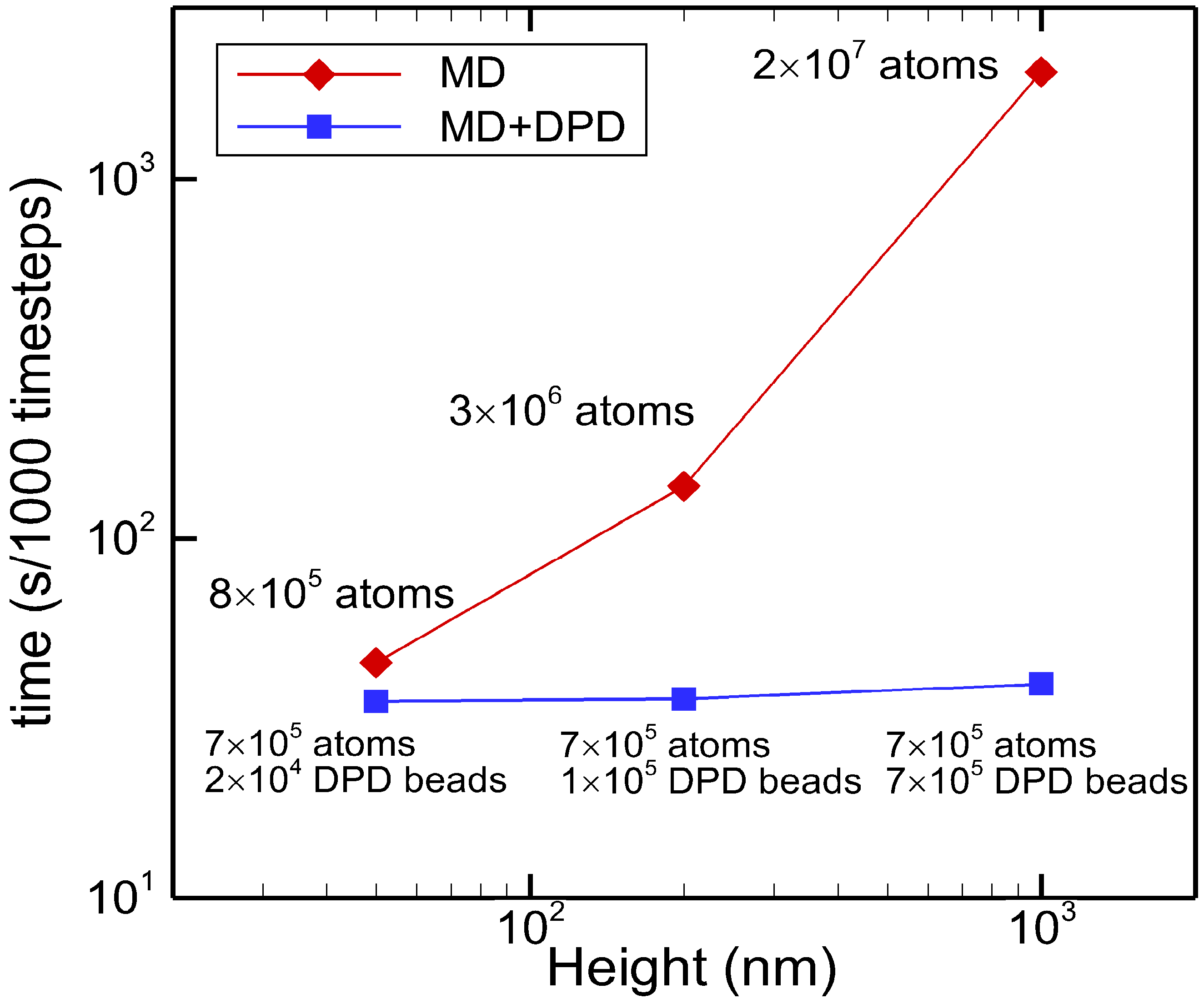}
\caption{Comparison of computational efficiency between MD and MD-DPD simulations for different channel heights. In the MD-DPD coupled simulations the MD subdomain remains fixed at $35~nm$ and we change the size of the DPD subdomain. All simulations were performed using 128 Intel Xeon E5-2670 CPUs. The timings were collected for $1\,000$ time steps for each of the simulations. }
\label{fig:efficiency}
\end{figure}
The advantage of the MD-DPD coupling method cannot be fully presented if the simulation system is relatively small. As we increase the size of system to $200~nm$ and $1~\mu m$ in height, the simulation efficiency for each case performed on a cluster of 128 CPUs is displayed in Fig.~\ref{fig:efficiency}. When the system size is increased from nano-meter to micro-meter, the computational cost of the coupled MD-DPD algorithm increases slightly, while the cost for a single MD simulation increases exponentially fast. For the height of $1~\mu m$, the computational cost of a MD simulation is two orders of magnitude higher than the MD-DPD simulation.

\section{Conclusions}\label{sec:4}
Bio-inspired hierarchical nanostructured surfaces play an important role in surface science for creating functional surfaces. A fundamental understanding of the relation between the surface structures and their functions can lead to a better and more effective design in diverse physical and biomedical applications.
The multiscale features originating from these hierarchical structures span across a wide range of spatio-temporal scales, and hence, are well beyond the capability of any single simulation method.
To this end, we developed a hybrid algorithm that couples molecular dynamics (MD) and dissipative particle dynamics (DPD) to solve the multiscale problems encountered in functionalized surfaces for nanoflow. The coupled method is able to cover different spatio-temporal scales and provide efficient simulation without losing local atomistic details.

We validated the coupled MD-DPD method using time-dependent simple flows such as Couette and Poiseuille flows. In both cases, we  observed smooth velocity and density profiles through the entire channel. Furthermore, the results show time-dependent accuracy in comparison with analytic solution. This MD-DPD coupled method can be used in many diverse applications in  physical and biological systems. Here, we used two examples to demonstrate the effectiveness of the MD-DPD method for multiscale systems.  For physical systems, we investigated the dynamics of polymer brushes under flow. In the MD domain, a detailed PDMS polymer model was established while the DPD domain included the solvent only. Our simulation results verify a previously established scaling law of slip length as a function of the imposed shear stress. For the biological system, we simulated flow over a surface grafted with a glycocalyx layer, fully immersed in the MD domain. In a test simulation with a simulation box of $50~nm$ in width, the MD-DPD method obtains identical velocity profiles with the MD alone simulation as well as a continuum-based numerical solution. Then we compare the simulation efficiency for the glycocalyx system with a height of $50~nm$, $200~nm$ and $1~\mu m$. As the simulation box grows in $z$-direction, the MD-DPD method performs stable simulations at very small extra cost while the MD simulation becomes prohibitively expensive.

\begin{figure}[t!]
\centering
\includegraphics[width=0.8\columnwidth]{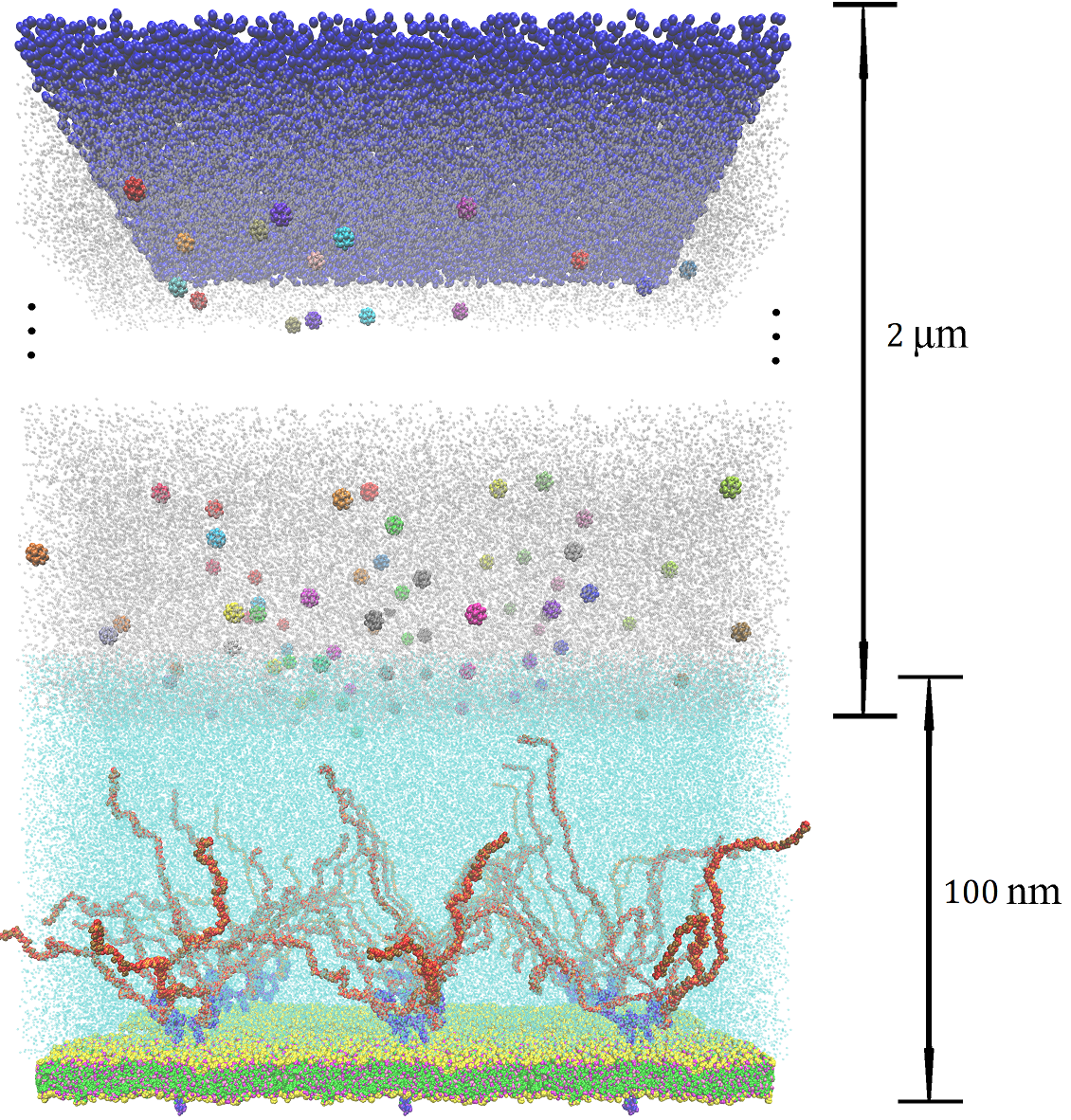}
\caption{Sketch of a setup for simulating the transport of drug delivering nanoparticles to the glycocalyx. The size of the DPD domain is typically $1$-$2~\mu m$ to represent the cell-free layer in arterioles. The nanoparticles are transported in the DPD solvent (plasma) and as they cross into the MD domain, they are endowed with molecular functionality.}
\label{fig:glycoillu}
\end{figure}

Another possible application of the MD-DPD method is to simulate accurately targeted drug delivery of functional nanoparticles (with size of tens of nanometers) to the specific site of the glycocalyx layer~\cite{2008De,2012Wang} (see Fig.~\ref{fig:glycoillu}). A typical human arteriole is about 50 microns, and the cell-free layer (CFL) is about 2 microns~\cite{2010Fedosov}. In the CFL region the solvent is plasma, which is similar to the DPD domain we employed in the glycocalyx MD-DPD simulation. The additional complexity, however, comes from the transport of nanoparticles through the CFL stream. These nanoparticles are functionalized with molecular level ligands and proteins, which somehow have to be resolved in order to correctly capture their interaction with the glycocalyx. However, since they are transported through the coarse-grained DPD domain, it is impossible to resolve molecular details in such large domain. What we envision, however, is simply the transport of nanoparticles through the DPD domain and their re-definition as they cross into the MD domain by endowing them with the functionalized ligands, since these information is available by the drug-delivery experts~\cite{2013Allen}.

In this paper, we demonstrated the application of MD-DPD multiscale simulation with domain decomposition in both physical (polymer-brush) and biological (glycocalyx) systems, which can be readily applied to other multiscale problems with complex interfaces requiring atomistic resolutions locally. Future works should consider extending this coupled MD-DPD method to multiphase flows involving suspensions and platelets moving through the overlapping interface between solvers. Fig.~\ref{fig:glycoillu} shows a sketch of the MD-DPD multiscale simulation applied to functional nanoparticles transported through the CFL region for targeted drug delivery, where the nanoparticles are endowed with molecular functionality as they cross into the MD domain from the DPD bulk domain.

\section*{Acknowledgements}
The work was supported by National Natural Science Foundation of China (No.~21878298, 11602133) and National Institutes of Health (NIH) Grants U01HL114476 and U01HL116323.\ Y.\ Wang acknowledges financial support from the Chinese Scholarship Council (CSC).\ This research was conducted using computational resources and services at the Center for Computation and Visualization, Brown University.

\footnotesize{
\providecommand*{\mcitethebibliography}{\thebibliography}
\csname @ifundefined\endcsname{endmcitethebibliography}
{\let\endmcitethebibliography\endthebibliography}{}

}

\end{document}